\documentclass[
 preprint,footinbib,
 amsmath,amssymb,
 aps,prl,showpacs, showkeys
]{revtex4-1}
\usepackage[utf8]{inputenc}
\usepackage{graphicx}
\DeclareGraphicsExtensions{.png}

\let\oldAA\AA
\renewcommand{\AA}{\text{\normalfont\oldAA}}

\begin{document}

\title{High-T$_c$ superconductivity in H$_3$S: Pressure effects on superconducting critical temperature and Cooper-Pairs Distribution Function}

\author{J. A. Camargo-Martínez}
\email{jcamargo@unitropico.edu.co}
\affiliation{Grupo de Investigación en Ciencias Básicas, Aplicación e Innovación - CIBAIN, UNITRÓPICO, Yopal, Casanare, Colombia}

\author{G. I. González-Pedreros}
\affiliation{Facultad de Ciencias Básicas y Aplicadas, Universidad Militar Nueva Granada, Bogotá, Colombia}

\author{R. Baquero}
\affiliation{Departamento de Física, CINVESTAV-IPN, Av. IPN 2508, 07360, CDMX, M\'exico}

\date{\today}

\begin{abstract}

We use first-principles calculations to study pressure effects on the vibrational and superconducting properties of H$_3$S in the cubic $Im\bar{3}m$ phase for the pressure range where the superconducting critical temperature (T$_c$) was measured (155–225 GPa). The pressure effects were incorporated using the Functional Derivative Method (FDM). In this paper, we present for the first time, the Cooper-Pairs Distribution Functions D$_{cp}$($\omega,T_c$) for H$_3$S, which will allow to identify the spectral regions where Cooper-Pairs formation at temperature T$_c$ is more favorable. We analyzed in detail the pressure effects on the electron-phonon spectral density function $\alpha^2$F$(\omega)$ and phonon density of states (PhDOS) and their relationship with T$_c$. The FDM manages to reproduce the trend of the pressure dependence of critical temperature, in good agreement with experimental data in the range of 155 to 190 GPa. The $D_{cp}(\omega, T_c)$ suggests that the low-frequency vibration region is where Cooper-Pairs are possible, which means that S-vibrations have an important role in the H$_3$S superconductivity properties.

\end{abstract}


\maketitle

\section{Introduction}

The effects of high pressures on superconducting properties of sulfur hydrides have been an important topic of study and discussion in the last years, even more so when one of them (H$_3$S) reached in 2015 the record of the experimental critical temperature (T$_c$), 203 K at 155 GPa reported by Drozdov et al.~\cite{Tc}. Experimentally, it was confirmed that H$_3$S superconductor behavior originates from electron-phonon interactions~\cite{EF}. In electron-phonon superconductors, pressure affects the vibration spectrum and electron-phonon interaction by shifting it to higher frequencies which can enhance or lower the critical temperature depending on details of the system under study~\cite{San}. This has been put in evidence by several experimental works~\cite{2a,3a,4a,5a,6a,7a,8a,9a,10a}. In H$_3$S, T$_c$ decreases when the pressure is increased~\cite{Tc}.

The relationship between T$_c$ and pressure in H$_3$S with cubic Im$\bar{3}$m structure has been explored theoretically in the framework of the Migdal–Eliashberg theory (MET) using different approximations; the Allen–Dynes equation~\cite{Tc8}, Allen-Dynes-modified McMillan equation with self-consistent harmonic and anharmonic approximations (SSCHA)~\cite{Tc15,Tc4}, isotropic Migdal-Eliashberg equations~\cite{Tc8} for stable isotopes~\cite{Tc12} and density functional theory for superconductors (SCDFT)~\cite{Tc6} including screened Coulomb repulsion via the random-phase approximation (RPA)~\cite{Tc5}. These theoretical results in some cases seem to reproduce experimental ones, however, there is no consensus on the method to estimate both the tendency and T$_c$ values measured.

From electronic and vibrational spectral properties obtained by first-principles calculations for a superconductor material, and taking into account the nature of Cooper-Pairs, González-Pedreros et al.~\cite{San2} recently reported a novel method to determine a Cooper-Pairs distribution function D$_{cp}$($\omega,T_c$), which seeks to find the vibrational energy ($\omega$) regions where Cooper-Pairs formation at temperature T$_c$ is more favorable. The D$_{cp}$($\omega,T_c$) function is given by
\begin{equation}\label{Dcp}
D_{cp}(\omega,T_c)=\int_{E_F-\omega_s}^{E_F+\omega_s}\int_{E_F-\omega_s}^{E_F+\omega_s} g_{ep}^s (\epsilon,\omega,T_c)\times g_{ep}^b (\epsilon'+\omega,\omega,T_c)\times \alpha^2(\omega)d\epsilon d\epsilon'.
\end{equation}

Here, $g_{ep}^s (\epsilon,\omega,T_c) \times g_{ep}^b(\epsilon'+\omega,\omega,T_c)$ is the probability at T$_c$ that: i) one electron is in energy state $\epsilon$, a second one is in energy state $\epsilon' + \omega$, ii) two empty electronic energy states  $\epsilon + \omega$ and $\epsilon'$, iii) two electrons are coupled to a phonon with energy $\omega$, iv) a vibrational energy state $\omega$,  v) an additional vibrational energy state $\omega$, and vi)  the electrons coupling with a phonon,  $\alpha^2(\omega)$. The calculation contains the contribution of all the electrons in the energy interval $\pm\omega_s$ around the Fermi level ($E_F\pm\omega_s$). For more details, see ref.~\cite{San2}.

In this paper, we report a theoretical study of pressure effects on the conventional superconductor H$_3$S by first-principles calculations for the pressure range where the T$_c$ was measured (155–225 GPa). The effects of pressure on T$_c$ are incorporated using the Functional Derivative Method (FDM)~\cite{San}. The D$_{cp}$($\omega,T_c$) is determined following the procedure proposed by González-Pedreros et al.~\cite{San2}. On the overall, our results agree very well with the experiment tendency and the calculated D$_{cp}$($\omega,T_c$) allowed us to identify the vibrational energy intervals where for H$_3$S the Cooper-pairs are possible.  

\section{Method of Calculation}  

The lattice dynamics, phonon densities of states (PhDOS) and the Eliashberg function $\alpha^2F(\omega)$ are calculated using the first-principle pseudopotential plane-wave method based on the density functional theory and the density functional perturbation theory as implemented in the Quantum-Espresso package~\cite{QS}. We used 70 Ry cut-off for the plane-wave basis. To integrate over the Brillouin zone (BZ), we used for the electronic integration a \textbf{k}-grid of $24\times 24\times 24$ and to compute phonon frequencies a \textbf{q}-grid of $8\times 8\times 8$ according to the Monkhorst–Pack scheme~\cite{MP}. We adopted the Vanderbilt ultrasoft pseudopotential~\cite{V} and a generalized gradient approximation (GGA) of the Perdew–Burke–Ernzerhof type (PBE) for the exchange-correlation energy functional~\cite{PBE}. In this work, the pressure effects are performed in the range where the high-T$_c$ was measured (155–225 GPa)~\cite{Tc}. 

The stable structures of the $Im\bar{3}m$ H$_3$S for each pressure are obtained relaxing the internal and external degrees of freedom using the Broyden–Fletcher–Goldfarb–Shanno (BFGS) quasi-Newton algorithm. This body-centered cubic structure, which reveals the superconducting phase in pressure range 155–225 GPa, was confirmed by X-ray diffraction experiment (150 GPa)~\cite{Nexpt}. The Coulomb pseudopotential $\mu^*$ at each pressure was calculated following the procedure suggested by Daams et al.~\cite{DA}.

The effects of pressure on T$_c$ were are calculated by means of the Functional Derivative Method (FDM)~\cite{San} (based on ideas from Rainer and Bergman~\cite{DF}) calculating $\Delta T_c$ from

\begin{equation}\label{Tc}
\Delta T_c =  \int_0^{\infty}{ \left.\frac{\delta T_c}{\delta \alpha^2F(\omega)} \right|_{P_i} \Delta \alpha^2F(\omega)},  
\end{equation}

where $\Delta \alpha^2F(\omega)=\alpha^2F(\omega,P_{i+1})-\alpha^2F(\omega,P_i)$. For more details, see ref.~\cite{San}

\section{Results and discussion} 

\begin{figure}[!ht]
\centering
\includegraphics[width=0.8\textwidth]{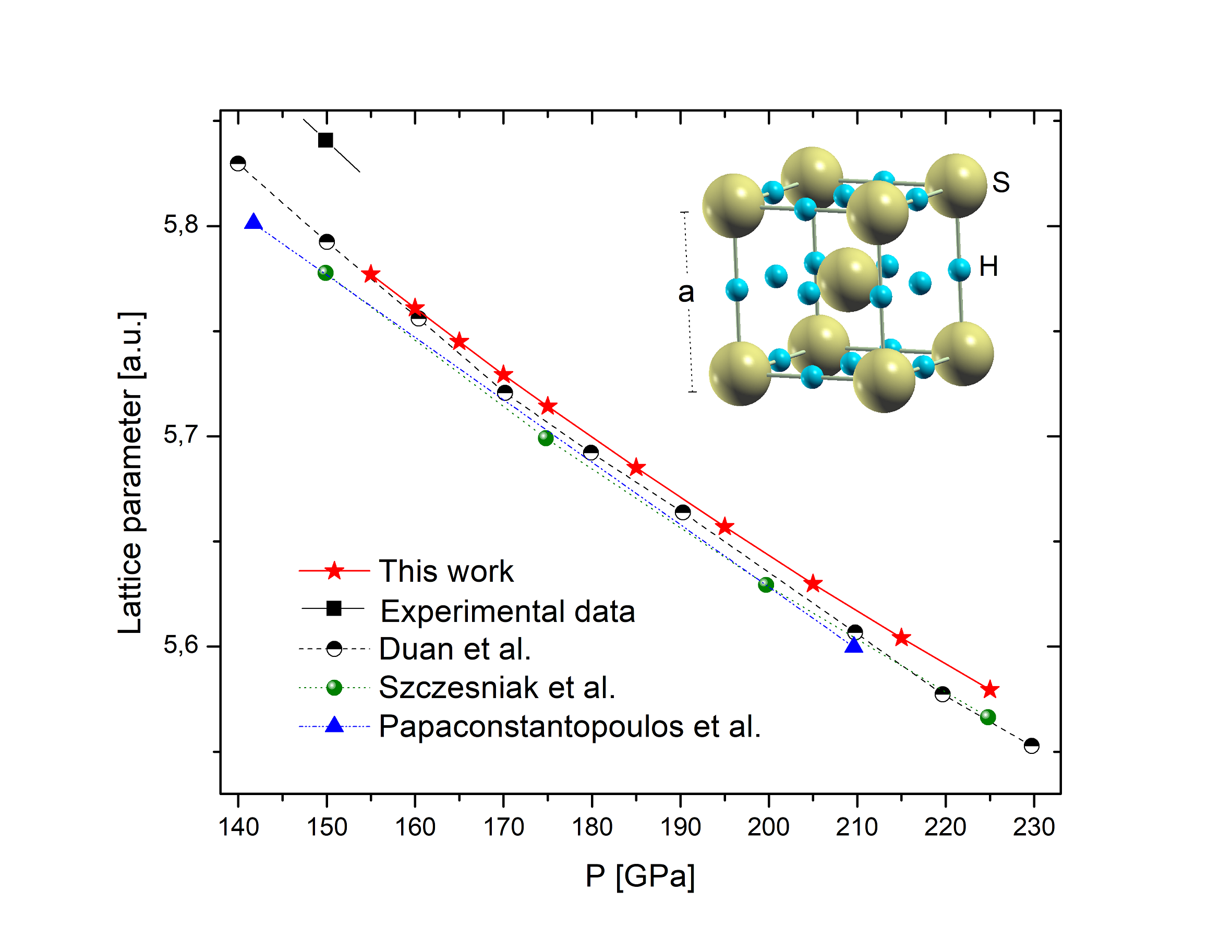} 
\caption{(Colour on-line) The calculated lattice parameters as a function of pressure and their comparison with other theoretical predictions~\cite{Tc12,Tc13,Tc17}. The experimental data was reported by Einaga et al.~\cite{Nexpt}. Inset presents the $Im\bar{3}m$ structure of H$_3$S (obtained by XCrySDen~\cite{Xc}).}
\label{LP} 
\end{figure}

The calculated lattice parameters as a function of pressure and their comparison with other theoretical predictions~\cite{Tc12,Tc13,Tc17} are shown in Fig.~\ref{LP}. It is observed that our results agree well with previous theoretical reports. In general, these predictions have a different around 0.06 $\AA$ under the experimental value at 150 GPa~\cite{Nexpt}. A fit of the Birch–Murnaghan equation (BME)~\cite{BM} to our data points gives the zero-pressure bulk modulus $B_0 = 85.76$ GPa and zero-pressure derivative of the bulk modulus $B'_0 = 3.9$, which are in good agreement with the values fitted to the experimental data reported by Errea et al~\cite{Tc15}.

\begin{figure}[!ht]
\centering
\includegraphics[width=0.8\textwidth]{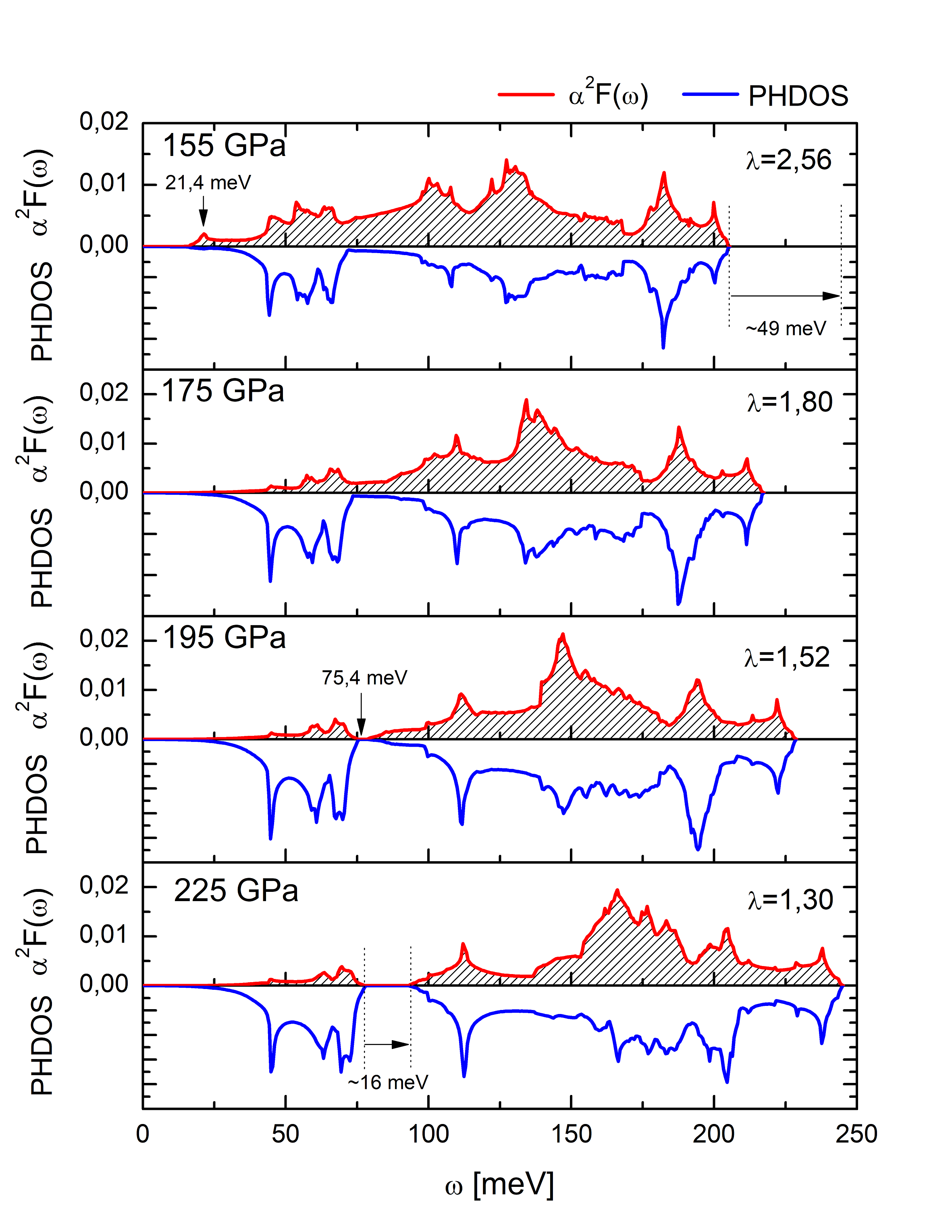} 
\caption{(Colour on-line) Eliashberg spectral function $\alpha^2F(\omega)$ and the phonon density of states (PhDOS) for H$_3$S calculated at different pressures. Note the electron-phonon coupling constants ($\lambda$) determined for each spectrum.}
\label{F2} 
\end{figure}

The Eliashberg spectral function $\alpha^2$F$(\omega)$ and the phonon density of states (PhDOS) calculated at different pressures are shown in Fig.~\ref{F2}. We do not observe any imaginary frequency vibrations in our calculation, which confirms the dynamic stability of $Im\bar{3}m$ structure in this pressure range.

It is observed that pressure induces a progressive decrease of the area under the $\alpha^2F(\omega)$ spectrum mainly in the range of 10 to 90 meV, and an almost rigid displacement of the $\alpha^2F(\omega)$ and PhDOS spectrum toward higher frequencies (hardening), from frequencies greater than 75.4 meV. At 195 GPa the $\alpha^2F(\omega)$ and PhDOS spectrum show a significant decrease in their spectral contributions around 75.4 meV, which entails a gap ($\sim$16 meV) in both spectrum at 225 GPa. This gap implies the reduction of the contribution of vibrational modes on the electron-phonon coupling constant ($\lambda $) and therefore, effect on T$_c$.

The total shifting of the $\alpha^2F(\omega)$ and PhDOS spectrum induced by pressure (155 to 225 GPa) reaches $\sim$49 meV. However, this shifting at frequencies below 75 meV is not greater than $\sim$3.0 meV. The $\alpha^2F(\omega)$ spectrum at 155 GPa reveals a small peak at 21.4 meV, which is not observed at higher pressures both in this and in other works. Nevertheless, a similar peak can be observed in calculations using the stable S isotopes $^{33}$S and $^{34}$S at 155 GPa~\cite{Tc12}. So, this peak seems to be related to a specific vibrational frequency of S atom which disappears with increasing pressure. Although it does not show a great intensity, it is a particular behavior of low-frequency vibrations region at 155 GPa which ends up contributing to high T$_c$. We do not find other theoretical results at 155 GPa that allow us to compare this behavior except for the one already mentioned.

On the other hand, it is observed that the increase in pressure induces a decrease in $\lambda$. It is important to note that a high value of $\lambda$ implies a high value of T$_c$.

\begin{figure}[!ht]
\centering
\includegraphics[width=0.9\textwidth]{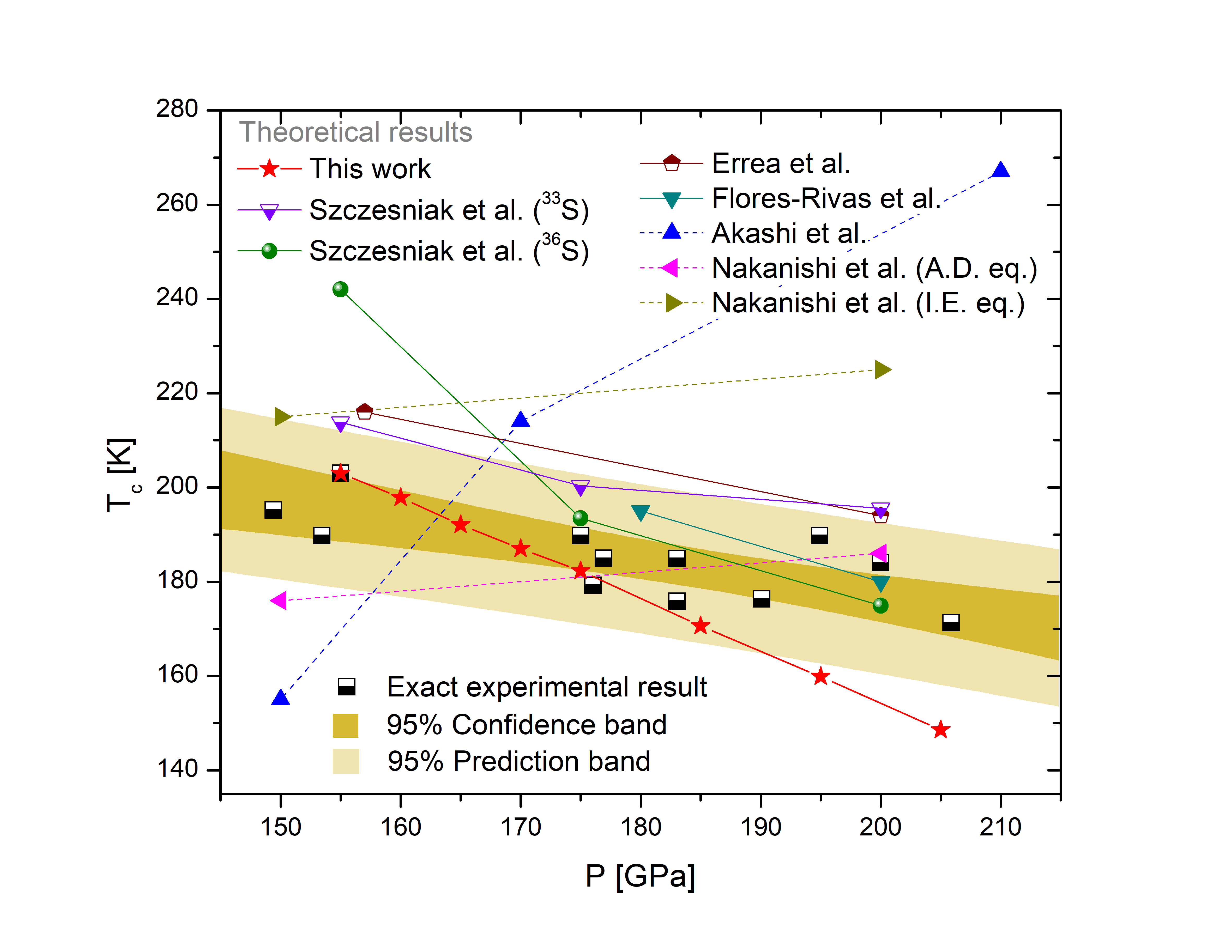} 
\caption{(Colour on-line) Effect of pressure on the critical temperature (T$_c$) measured~\cite{Tc} and calculated~\cite{Tc12,Tc5,Tc15,Tc4,Tc6,Tc8} for H$_3$S. Red stars symbols correspond to T$_c$'s calculated in this work using the Functional Derivative Method (FDM)~\cite{San}. Experimental T$_c$ value at 155 is the starting data in implementation of the FDM. The dash lines indicate the disagreement with the experimental tendency reported. Confidence and prediction bands were obtained by a linear fit to the experimental data.}
\label{F1} 
\end{figure}

Fig.~\ref{F1} shows the pressure dependence of the superconducting critical temperature (T$_c$) measured~\cite{Tc} and calculated in this work using the FDM, and its comparison with previous theoretical reports~\cite{Tc12,Tc5,Tc15,Tc4,Tc6,Tc8}. On the overall, the FDM manages to reproduce the trend established by the experimental results reported by Drozdov et al.~\cite{Tc}. Our calculations show a $dT_c/dP=-1.0$ K/GPa which is slightly greater than reported by the experiment ($dT_c/dP=-0.4$). The negative sign implies the reduction of {\it adequate physical conditions} that induced a high T$_c$. The projected PhDOS reported in the literature for H$_3$S~\cite{Tc13,Tc12,Tc4,Tc1,Tc12} show that the low-frequency vibrations region (below 70 meV) is due mainly to the vibrations of the S atom. 
According to this and our calculations, we suggest two possible conditions that reduce T$_c$; the decrease in the spectral contribution of electron-phonon coupling at low-frequency vibrational modes (S atoms) and slight hardening of high-frequency vibration modes of H atoms. In relation to this, Amsler~\cite{Tc1} and Flores-Livas et al.~\cite{Tc5} reported that when Se atom is included in H$_3$S the contribution of atom S at PhDOS spectrum (in low-frequency vibration region) decreases and consequently T$_c$ decreases. It is evident that S atom play a relevant role in the high T$_c$ of H$_3$S.

Our T$_c$'s are in good agreement with experimental ones in the range of 155-190 GPa, lying within 95\% confidence band (155-175 GPa). For pressures greater than 190 GPa, our results are outside of the 95\% band prediction. We suggest that the discrepancy at high pressures could be due to no inclusion of anharmonic and screened Coulomb repulsion effects, which has shown acceptable T$_c$ values in other work~\cite{Tc15,Tc4,Tc5}. Concerning to other works, it is observed that the calculations reported by Szcześniak et al.~\cite{Tc12}, Flores-Livas et al.~\cite{Tc5} and Errea et al.~\cite{Tc4} show a $dT_c/dp\sim-0.4$ in excellent agreement with the experiment trend. However, some of their values are outside the experimental range. Errea et al. propose that the inclusion of anharmonic effects is crucial for explaining the measured T$_c$~\cite{Tc4}. However, the results reported by Nakanishi et al.~\cite{Tc8} which included these effects, did not reproduce the experimental trend. The T$_c$(p) calculations using SCDFT reported by Akashi et al.~\cite{Tc6} show the greatest discrepancies with the experiment both in values and trends. However, Flores-Livas et al.~\cite{Tc5} using the same method including screened Coulomb repulsion effects~\cite{Tc4} obtained good results in the range 180-200 GPa.

We note that obtaining a good agreement for a specific value of T$_c$ does not necessarily imply an adequate reproduction of the tendency experimentally observed, as is the case of the results reported by Nakanishi et al.~\cite{Tc8}, in which using the Allen-Dynes equation was calculated at 200 GPa a T$_c$ of 186 K with excellent agreement with the experiment, but their data reveal a positive $dT_c/dp$, in contradiction with the experimental behavior. They reported at 150 GPa a T$_c$ value of 19 K below the experimental one.

An important number of theoretical calculations of T$_c$ at 200 GPa have been reported in the literature, which oscillate between 180 and 284 K~\cite{Tc12,Tc5,Tc8,Tc13,Tc1,Tc3,Tc4,Tc10,Tc5,Tc7,Tc8,Tc9,Tc11,Tc14}, some of them~\cite{Tc12,Tc5,Tc8,Tc1,Tc10,Tc9} within the experimental range (160 - 192 K). In our case, the FDM determined a value (extrapolated) at 200 GPa of 155.4 K, which is 4.6 K below the experimental range. 

\begin{figure}[!ht]
\centering
\includegraphics[width=1.0\textwidth]{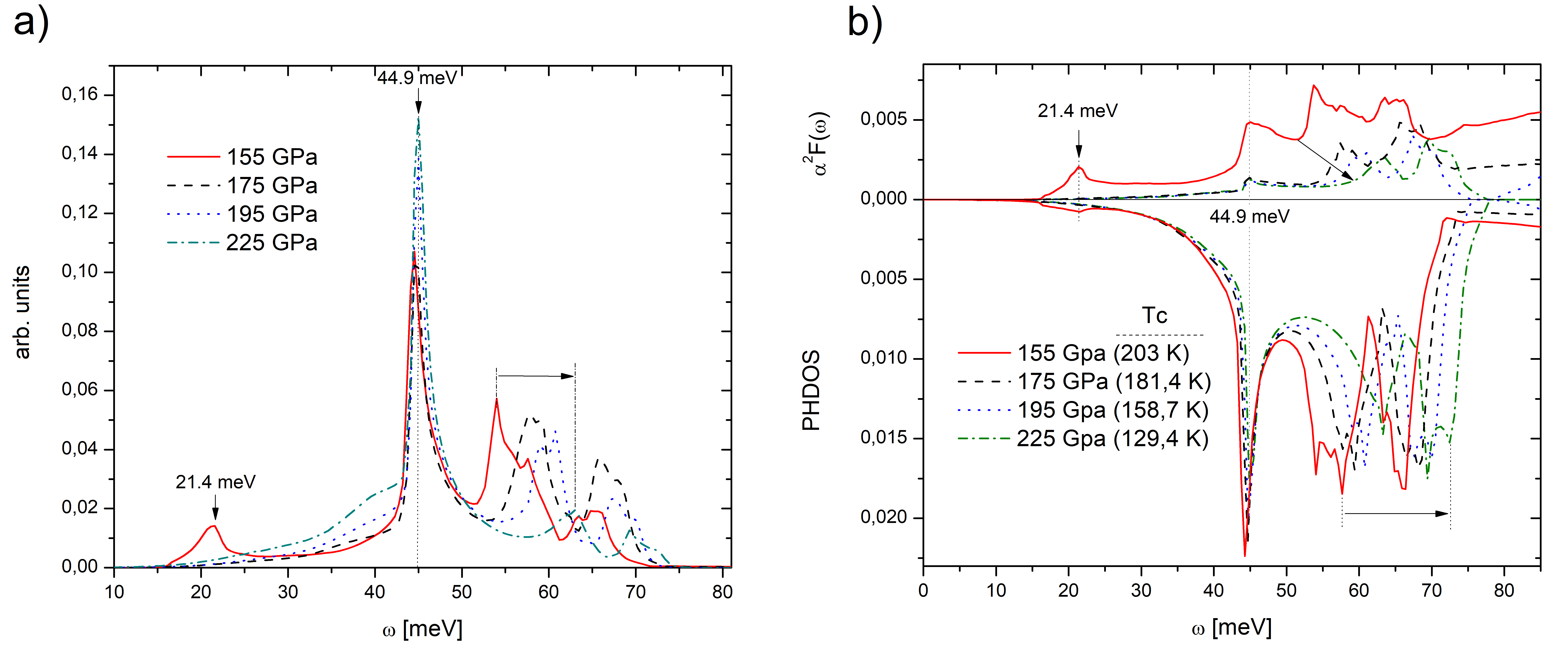} 
\caption{(Colour on-line) (a) The Cooper-Pairs Distribution Functions (D$_{cp}$($\omega,T_c$)) of H$_3$S at different pressures. (b) Zoom of Eliashberg spectral function $\alpha^2F(\omega)$ and the phonon density of states (PhDOS) for H$_3$S calculated at different pressures at range 0 - 85 meV (D$_{cp}$($\omega,T_c$) range). The arrows indicate the important changes in these spectrum. The T$_c$ values calculated associated to these spectrums are specified.} 
\label{Dcp1}
\end{figure}

In Fig.~\ref{Dcp1}(a) the Cooper-Pairs Distribution Functions D$_{cp}$($\omega,T_c$) of H$_3$S calculated at different pressures are presented. In general, it is observed that the system reveals the existence of possible Cooper-Pairs only in the frequency range from 10 to 80 meV, which is a surprising result since it significantly limits the analysis of the contribution of higher frequencies in the $\alpha^2 F(\omega)$ and PhDOS spectrum in the interpretation of the physical mechanisms that induce the superconducting state and respective T$_c$. With respect to the effects of pressure, for frequencies higher than $\sim$44 meV the D$_{cp}$($\omega,T_c$) show a shifting (not rigid) to the right and progressive decrease of the area under the curve of the respective D$_{cp}$($\omega,T_c$) function.

As seen above in the PhDOS spectrum (see Fig.~\ref{F2}) the D$_{cp}$($\omega,T_c$) functions show an important contribution at 44.9 meV which, due to the effects of pressure it is weakly displaced ($\sim$ 1 meV) towards higher energies. As in the $\alpha^2F(\omega)$ spectrum, only the D$_{cp}$($\omega,T_c$) calculated at 155 GPa reveals a peak at 21.4 meV. The contribution of this peak could be one of the differentiating factors that induce in the H$_3$S the mechanisms to reveal the highest T$_c$. At the range 25 - 44 meV  an increase in area under the curve is observed, as opposed to the decrease in T$_c$, with the increase in pressure.

In Fig.~\ref{Dcp1}(b) the zoom at the range 0 - 85 meV for $\alpha^2F(\omega)$ and PhDOS spectrum calculated at different pressures are shown. Both $\alpha^2F(\omega)$ y PhDOS spectrum reveal a peak at 21.4 meV only at 155 GPa. It is evident that the pressure induces a considerable decrease of the area under the $\alpha^2F(\omega)$ spectrum with respect to the one at 155 GPa. This decrease can be directly related to the decrease in T$_c$. At energies greater than 50 meV the PhDOS show a quasi-rigid shift to the right.

Electrons in a Cooper pair demand occupied and vacant states near Fermi level and by electron-phonon interaction, they move between these states. The energy proximity between S-s, S-p and H-s orbitals in H$_3$S~\cite{BER} and their contribution to Fermi level~\cite{Tc12} are part of the conditions that electrons require in a Cooper pair. D$_{cp}$ functions for H$_3$S show that Cooper pair phonon energy lies in the 10- to 80-meV interval (D$_{cp}$ region), and the main phonon energy contributes to Cooper pair formation is at $\omega_{cp} \sim44.9$ meV in all pressures. According to pressure increases, the $\alpha^2F(\omega)$ spectrum (in D$_{cp}$ region) decreases significantly in intensity and moves slightly to higher frequencies, then the conditions to Cooper pair formation weakens and $T_c$ falls.

The Projected PhDOS reported~\cite{Tc14,Tc4,Tc12} below 70 meV show the vibrational modes both of S and H atoms (see Fig. 4 in ref.~\cite{Tc12}). This brings to think that H$_3$S superconductivity is established by electrons of S-s, S-p and H-s orbitals that interact with phonons in the D$_{cp}$ region, where S vibrations mainly contribute, however H vibrations has an important participation. The physical mechanism underlying the high-temperature superconductivity of H$_3$S with pressure is metallization of covalent bond~\cite{BER}, however, when the pressure is increased electron-electron interactions become more strong whereby $T_c$ decreases. 

\begin{figure}[!ht]
\centering
\includegraphics[width=1\textwidth]{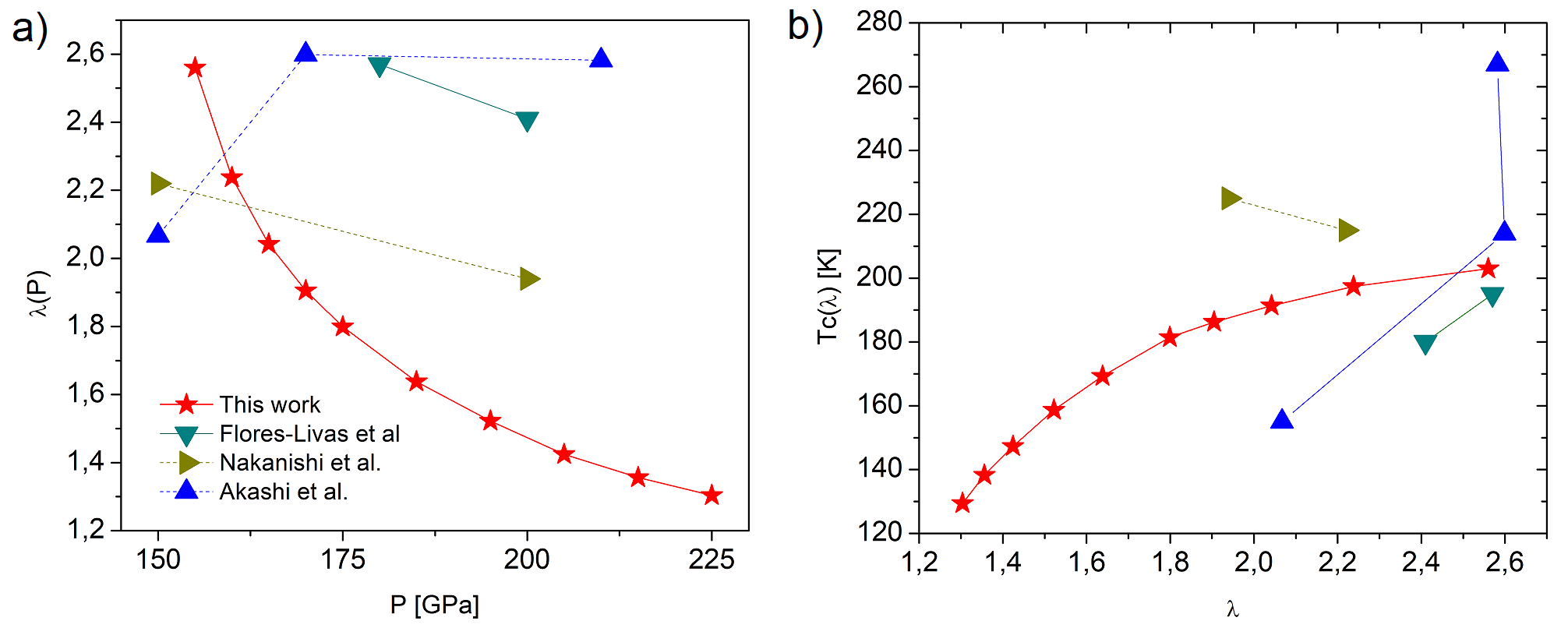} 
\caption{(Colour on-line) a) Electron phonon coupling constant $\lambda$ as a function of pressure. b) Critical temperature T$_c$ as a function of the electron phonon coupling $\lambda$. The comparison with other works~\cite{Tc12,Tc5,Tc6,Tc8} is included in each case.} 
\label{lm} 
\end{figure}

In Fig.~\ref{lm}, the electron-phonon coupling constant $\lambda$ as a function of pressure, and T$_c$ as a function of $\lambda$ are presented in Fig.~\ref{lm}.

It is observed in Fig.~\ref{lm}a that increase of pressure induces a non-linear decrease of $\lambda$, which is more significant in the pressure range 155-175 GPa where the decrease of the area under the $\alpha^2F(\omega)$ spectrum at low frequencies (associated mainly to S atom) is considerably greater (see Fig.~\ref{F2}). Flores-Livas et al.~\cite{Tc5} and Nakanishi et al.~\cite{Tc8} show a similar tendency but $\lambda$ does not decay so quickly with increasing P, as in our case. The $\lambda$ values reported by Akashi et al.~\cite{Tc6} show an opposite behavior compared to our results. We observed, in most cases, T$_c$($\lambda$) reveals that a strong $\lambda$ induces a high T$_c$ (Fig.~\ref{lm}b). A $\lambda$ value around 2.6 leads in a high T$_c$ for H$_3$S. The correlation between T$_c$ and $\lambda$ is in total according to the Migdal-Eliashberg theory~\cite{MC}.

\section{Conclusions}

In this work, we present a theoretical study of pressure effects on high-T$_c$ superconductivity in the conventional superconductor H$_3$S ($Im\bar{3}m$). The effects of pressure on T$_c$ were incorporated using the Functional Derivative Method (FDM). We present for the first time the Cooper-Pairs Distribution Functions (D$_{cp}$($\omega,T_c$)) calculated for H$_3$S, following the procedure proposed by González-Pedreros et al.~\cite{San2}.

We found that pressure induces a decrease of the area under the $\alpha^2F(\omega)$ spectrum mainly in the range of 10 to 90 meV. From frequencies greater than 75.4 meV an almost rigid displacement of the $\alpha^2F(\omega)$ spectrum and phonon density of states (PhDOS) toward higher frequencies is also observed. A small peak at 21.4 meV related to vibrational frequencies of S atom is only observed at 155 GPa, which generate a little contributions to reach the high $T_c$ of 203 K. As a first idea, we suggest two possible conditions that reduce T$_c$ for pressure effects in H$_3$S; the decrease in the spectral contribution of electron-phonon coupling at low-frequency vibrational modes (S atoms) and slight hardening of high-frequency vibration modes of H atoms. S atom ends up playing an important role in the high T$_c$ reached at 155 GPa in H$_3$S.

The FDM manages to reproduce the trend in $T_c$(P) established by experimental results. Our $T_c$ calculations are a good agreement with experimental data in the range of 155-190 GPa. The discrepancy at high pressures (190-225 GPa) could be due to no inclusion of anharmonic and screened Coulomb repulsion effects, which has shown acceptable T$_c$ values in other work. This idea will be studied in our next research works.

The $D_{cp}(\omega, T_c)$'s reveal that the 10- to 80-meV energy interval (D$_{cp}$ region) is where Cooper-Pairs are possible. According to pressure increases the $\alpha^2F(\omega)$ spectrum (in D$_{cp}$ region) decreases significantly in intensity and moves slightly to higher frequencies, then the conditions to Cooper pair formation weakens and $T_c$ falls. We suggest that H$_3$S superconductivity is established by electrons of S-s, S-p and H-s orbitals that interact with phonons in the D$_{cp}$ region, where S vibrations mainly contribute, 
with the contribution of H vibrations.

Finally, the electron-phonon coupling constant $\lambda$ shows an important correlation with T$_c$, in total according to the Migdal-Eliashberg theory.

\section{Acknowledgments}

The authors acknowledge to the General Coordination of Information and Communication Technologies (CGSTIC) at CINVESTAV for providing HPC resources on the Hybrid Cluster Supercomputer Xiuhcoatl.

\end{document}